\documentclass[reprint,amsmath,amssymb,aps,prb,showpacs]{revtex4-1}

\usepackage{graphicx}
\usepackage{epsfig}
\usepackage{dcolumn}
\usepackage{bm}
\usepackage{hyperref}
\hypersetup{backref=true,
 pdfnewwindow=true, colorlinks=true,
 linkcolor=blue, anchorcolor=blue,
 citecolor=blue, filecolor=blue,
 menucolor=blue, urlcolor=blue}

\usepackage{color}

\begin{document}
\title{Computing topological invariants without inversion symmetry}
\author{Alexey A. Soluyanov}
\email{alexeys@physics.rutgers.edu}
\author{David Vanderbilt}
\email{dhv@physics.rutgers.edu}
\affiliation{Department of Physics and Astronomy, Rutgers University,
Piscataway, New Jersey 08854-0849, USA}
\date{\today}

\def\TR{{\cal T}}
\def\TT{T}

\begin{abstract}
We consider the problem of calculating the weak and strong topological
indices in noncentrosymmetric time-reversal ($\TR$) invariant
insulators.  In 2D we use a gauge corresponding to hybrid Wannier
functions that are maximally localized in one dimension.
Although this gauge is not smoothly defined on the
two-torus, it respects the $\TR$ symmetry of the
system and allows for a definition of the $\mathbb{Z}_2$ invariant in terms
of time-reversal polarization.
In 3D we apply the 2D approach to $\TR$-invariant planes.
We illustrate the method with first-principles
calculations on GeTe and on HgTe
under $[001]$ and $[111]$ strain. Our approach differs from ones
used previously for noncentrosymmetric materials
 and should be easier to implement in {\it ab
initio} code packages.
\end{abstract}
\maketitle

\marginparwidth 2.7in
\marginparsep 0.5in
\def\dvm#1{\marginpar{\small DV: #1}}
\def\asm#1{\marginpar{\small AS: #1}}
\def\scr{\scriptsize}

\def\r{{\bf r}}
\def\R{{\bf R}}
\def\k{{\bf k}}
\def\G{{\bf G}}
\def\z2{{\mathbb{Z}_2}}

\section{INTRODUCTION}

A series of theoretical developments starting in 2005, showing that
non-magnetic insulators admit a topological $\z2$ classification in
two dimensions (2D) \cite{Kane-PRL05-a,Kane-PRL05-b} 
and then in three dimensions (3D),\cite{Fu-PRL07,Moore-PRB07}
has sparked enormous interest, especially after numerous 
realizations of such systems were confirmed both 
theoretically\cite{Bernevig-Science06,Fu-PRB07,Zhang-NatPhys09, 
Chadov-NatMat10, Klintenberg-arx10}
and experimentally.\cite{Konig-Science07,
Hsieh-Nature08,Hsieh-Science09,Xia-NatPhys09,Chen-Science09}
These developments, nicely summarized in some recent
reviews,\cite{Hasan-RMP10,Qi-PT10,Qi-arx10} have essentially given rise to
a new subfield of condensed-matter physics, with the topology
of the band structure now regarded as a fundamental characteristic of
the electronic ground state for semiconductors and insulators.  

The $\z2$ classification
divides  time-reversal ($\TR$) invariant band insulators into two classes:
ordinary ($\z2$-even) insulators that can be adiabatically converted to
the vacuum (or to each other) without a bulk gap closure, and
``topological'' ($\z2$-odd) ones that cannot be so connected
(although they can be adiabatically connected to each other).
Even and odd phases
are separated by a topological phase transition, and the bulk gap
has to vanish at the transition point, at least in a non-interacting
system.\cite{Schnyder-PRB08, Kitaev-AIP09}
The $\z2$-odd states are characterized by the presence of an odd
number of Kramers pairs of counterpropagating edge states in 2D,
or by an odd number of Fermi loops enclosing certain high-symmetry
points of the surface band structure in 3D.

In view of all this, there is an obvious motivation to develop simple
yet effective methods for computing the topological indices of
a given material.  For centrosymmetric crystals, a convenient
method was introduced in Ref.~\onlinecite{Fu-PRB07}, where
it was shown that the knowledge of the parity eigenvalues of the
electronic states at only four $\TR$-invariant momenta in 2D (or
eight of them in 3D) is sufficient to compute the topological
characteristics of a given material.  This approach is
limited to centrosymmetric systems, however, and the calculation
of the $\z2$ invariant for noncentrosymmetric insulators is not
so trivial.

One possible approach, suggested in Ref.~\onlinecite{Fukui-JPSJ07},
is based on the existence of a topological obstruction to choosing
a smooth gauge that respects the $\TR$ symmetry in the $\z2$-odd
case. For the implementation of this method, a gauge must be
chosen on the boundary of half of the Brillouin zone (BZ) in
such a way as to respect $\TR$ symmetry, which involves acting
with the time-reversal operator on one of the states from each
Kramers pair to construct the other. Although this method has been
implemented in the {\it ab initio} framework\cite{Xiao-PRL10,
Feng-PRL11, Wada-PRB11}, its implementation is basis-set dependent and involves
the application of a unitary rotation to the computed eigenvectors
when fixing the gauge, which may be tedious when there are many
occupied bands and basis states.

Another existing method\cite{Zhang-NatPhys09} relies on the
fact that the system will necessarily be in the $\z2$-even (normal)
state in the absence of spin-orbit (SO) coupling.  In this method, the
strength of the SO coupling is artificially tuned from $\lambda_{SO}=0$
(no SO coupling) to $\lambda_{SO}=1$ (full SO coupling), and a
closure of the band gap at some intermediate coupling strength is
taken as evidence of an inverted band structure.  However, a closure
of the band gap in the course of tuning  $\lambda_{SO}$ to full
strength is a necessary, but not a sufficient, condition for a
topological phase transition.  Therefore, in order to determine
whether the system is really in the topologically nontrivial phase,
a first-principles calculation of the surface states is carried out
in order to count the number of Dirac cones at the surface
of the candidate material. Such a calculation, although
illustrative, is quite demanding in terms of computational resources.

In summary, existing methods have some shortcomings, and it would be
very useful to develop a simple and effective method that would use
the electronic wavefunctions, as obtained directly from the
diagonalization procedure, to determine the desired topological indices.

In this paper we develop a method for computing $\z2$ invariants
that meets these criteria, and which is easy to implement in
the context of {\it ab initio} code packages. The method is based
on the concept of time-reversal polarization\cite{Fu-PRB06}
(TRP), but implemented in such a way that a visual inspection of
plotted curves is not required in order to obtain the topological
indices.  Instead, all the indices can be obtained directly as a
result of an automated calculation.  We describe the method, and
then verify it using centrosymmetric Bi and Bi$_2$Se$_3$ as
illustrative test examples before applying it to the more difficult
cases of noncentrosymmetric GeTe and strained HgTe.

The paper is organized as follows. In Sec.~\ref{sec:wcc} we 
start by reviewing the formalism
of TRP in the context of the $\z2$ spin pump in one dimension (1D),
emphasizing its relation to the charge centers of Wannier functions.
We then discuss the numerical implementation of these ideas 
to 2D and 3D cases, and suggest a simple numerical procedure
for calculating the $\z2$ invariant in noncentrosymmetric $\TR$-invariant
systems in Sec.~\ref{Sec:3}.  We further illustrate this method with {\it ab
initio} calculations in Sec.~\ref{sec:applic}, and present some
concluding remarks in Sec.~\ref{sec:concl}.

\section{${\z2}$ invariant via Wannier charge centers}
\label{sec:wcc}

In this section we review the notion of TRP and the definition of the
$\z2$ invariant in terms of TRP derived in Ref.~\onlinecite{Fu-PRB06}.
The definition arises by virtue of an analogy between a 2D $\TR$-invariant
insulator and a $\TR$-symmetric pumping process in a 1D insulator.
We further reformulate this definition in terms of Wannier
charge centers, setting the stage for the numerical method
discussed in the next section.

\subsection{Review of time reversal polarization}

Fu and Kane\cite{Fu-PRB06} considered a family of
1D bulk-gapped Hamiltonians $H(x)$ parametrized by a cyclic
parameter $t$ (i.e., $H[t+\TT]=H[t]$) subject to the constraint
\begin{equation}
  H[-t]=\theta H[t] \theta^{-1},
\label{constr}
\end{equation}
where $\theta$ is the time-reversal operator.  This can be
understood as an adiabatic pumping cycle, with $t$ playing the role
of time or pumping parameter.  The constraint of Eq.~(\ref{constr})
guarantees that the Hamiltonian $H(x)$ is $\TR$-invariant at the
points $t=0$ and $t=\TT/2$, while the $\TR$ symmetry is broken
at intermediate parameter values. 
If we also limit ourselves to
Hamiltonians having unit period, so that $H$ is invariant under
$x\rightarrow x+1$, then the
eigenstates may be represented by the periodic parts $|u_{nk}\rangle
=e^{-ikx}|\psi_{nk}\rangle$ of the Bloch states $|\psi_{nk}\rangle$.
At $t=0$ and $t=\TT/2$ the Hamiltonian is time-reversal invariant
and the eigenstates come in Kramers pairs, being degenerate at
$k=0$ and $k=\pi$.

Since the system is periodic in both $k$ and $t$, the
$|u_{nk}\rangle$ functions are defined on a torus. Moreover, the
system must also be physically invariant under a gauge transformation of
the form
\begin{equation}
|\tilde{u}_{nk}\rangle = \sum_m U_{mn}|u_{mk}\rangle
\label{gauge_tr}
\end{equation}
where $U(k,t)$ expresses the ${\cal U}({\cal N})$ gauge freedom to choose
${\cal N}$ representatives of the occupied space at each $(k,t)$.
We adopt a gauge that is continuous on the half-torus $t\in[0,\TT/2]$
and that respects $\TR$ symmetry at $t=0$ and $\TT/2$ in the sense
of Fu and Kane,\cite{Fu-PRB06} i.e.,
\begin{eqnarray}
|u^I_{\alpha,-k}\rangle &=& - e^{i\chi_{\alpha k}}\theta|u^{II}_{\alpha k}\rangle, \nonumber\\
|u^{II}_{\alpha,-k}\rangle &=& e^{i\chi_{\alpha,-k}}\theta|u^I_{\alpha k}\rangle.
\label{kramers}
\end{eqnarray}
Here the occupied states $n=1,...,{\cal N}$ have been relabeled in
terms of pairs $\alpha=1,...,{\cal N}/2$ and elements $I$ and $II$
within each pair.
Note that Eq.~(\ref{kramers}) is a property which is not preserved
by an arbitrary ${\cal U}({\cal N})$ transformation.  It allows
the Berry connection
\begin{equation}
 {\cal A}(k)=i\sum_n \langle u_{nk}| \partial_k |u_{nk} \rangle
\end{equation}
to be decomposed as
\begin{equation}
{\cal A}(k)={\cal A}^I(k)+{\cal A}^{II}(k)
\end{equation}
where
\begin{equation}
 {\cal A}^S(k)=i\sum_\alpha \langle u^S_{\alpha k} |\partial_k| u^S_{\alpha k} \rangle
\end{equation}
and $S=I,II$.  Having chosen a gauge that obeys these conventions at
$t=0$ and $\TT/2$ and evolves smoothly for intermediate $t$,
\footnote{Since we do not constrain the gauge of the 1D system to
obey any particular symmetries at intermediate $t$, it is always
possible to perform the unitary mixing at intermediate $t$ in such
a way that the pair of ``bands'' belonging to the same $\alpha$
at $t$=0 also belong to the same $\alpha$ at $t$=$\TT/2$.}
the ``partial polarizations''\cite{Fu-PRB06}
\begin{equation}
P_\rho^{S}=\frac{1}{2\pi}\oint dk {\cal A}^S(k)
\label{partial}
\end{equation}
can be defined such that their sum is the total charge
polarization\cite{King-Smith-PRB93}
\begin{equation}
P_\rho=\frac{1}{2\pi}\oint dk {\cal A}(k)=P^I_\rho + P^{II}_\rho.
\label{ch_polar}
\end{equation}

Note that the total polarization is defined only modulo an integer
(the quantum of polarization) under a general $U({\cal N})$ gauge
transformation, while the ``partial polarization" is not gauge invariant
at all. A quantity that {\it is} gauge-invariant is the change in total
polarization during the cyclic adiabatic evolution of the Hamiltonian, and
using Eq.~(\ref{constr}) it follows that
\begin{equation}
P_\rho(\TT)-P_\rho(0)=C
\label{Chern}
\end{equation}
where $C$ is the first Chern number, an integer topological invariant
corresponding to the number of electrons
pumped through the system in one cycle of the pumping
process.\cite{Thouless-PRL82} For a $\TR$-invariant pump
that satisfies the conditions of Eq.~(\ref{constr}), $C$ must be zero.

In order to describe the $\z2$ invariant of a $\TR$-symmetric system
in a similar fashion, the ``time reversal polarization" was introduced
as\cite{Fu-PRB06}
\begin{equation}
P_\theta = P_\rho^{I}-P_\rho^{II}.
\end{equation}
Then the integer $\z2$ invariant can be written as
\begin{equation}
\Delta=P_\theta(\TT/2)-P_\theta(0) \mod 2.
\label{z2}
\end{equation}

To summarize, the $\z2$ invariant is well defined via
Eq.~(\ref{z2}) when the gauge respects $\TR$-symmetry at $t=0$ and $\TT/2$
and is continuous on the torus between these two parameter values.
Note, however, that while such a gauge choice is possible on the
half-torus even for the $\z2$-odd case ($\Delta$=1), it can only
be extended to cover the full torus continuously in the
$\z2$-even case ($\Delta$=0).\cite{Fu-PRB06, Roy-PRB09-a, Soluyanov-PRB11}

\subsection{Formulation in terms of Wannier charge centers}
\label{sec:formulation}

Let us now rewrite Eq.~(\ref{z2}) in terms of the Wannier charge
centers (WCCs).
By definition the Wannier functions (WFs) belonging to
unit cell $R$ are
\begin{equation}
|Rn\rangle =\frac{1}{2\pi} \int_{-\pi}^{\pi} dk e^{-ik(R-x)}|u_{nk}\rangle.
\label{wfdef}
\end{equation}
The WCC $\bar{x}_n$ is defined as the expectation value
$\bar{x}_n=\langle 0n| \hat{X}| 0n \rangle$ of the position operator
$\hat{x}$ in the state $| 0n \rangle$ corresponding to one of the WFs
in the home unit cell $R=0$. Equivalently,
\cite{Zak-PRL89, King-Smith-PRB93}
\begin{equation}
\bar{x}_n=\frac{i}{2\pi}\int_{-\pi}^{\pi}dk \langle u_{nk}
|\partial_k|u_{nk}\rangle.
\label{defwcc}
\end{equation}
Except in the single-band case, the individual $\bar{x}_n$ are
{\it not} independent of a general gauge transformation as in
Eq.~(\ref{gauge_tr}).  However, the {\it sum} over all WCCs in the
unit cell {\it is} a gauge-independet quantity (modulo a lattice
vector, i.e., mod 1 in our notation).\cite{King-Smith-PRB93}
For the present purposes we adopt the gauge of Eq.~(\ref{kramers})
and construct WFs $|R\alpha,S\rangle$ by inserting
$|u_{\alpha k}^S\rangle$ into the definition of Eq.~(\ref{wfdef}).
In this gauge
\begin{equation}
\bar{x}_\alpha^{I}=\bar{x}_\alpha^{II} \mod 1,
\end{equation}
as follows from Eqs.~(\ref{kramers}) and
(\ref{defwcc}) and use of the continuity condition
$\chi_{\alpha,-\pi}=\chi_{\alpha,\pi}+2\pi m$, where $m$
is an integer.
Since we have also insisted
on the gauge being continuous for $t\in[0,\TT/2]$,
it is possible to follow the evolution of each
WCC during the half-cycle. Taking into account that
$\sum_\alpha \bar{x}_\alpha^s=(1/2\pi)\oint_{\rm BZ}{\cal A}^S$
for $S=I,II$, Eq.~(\ref{z2}) yields
\begin{equation}
\Delta = \sum_\alpha\left[ \bar{x}_\alpha^I(\TT/2)
- \bar{x}_\alpha^{II}(\TT/2)\right]
-\sum_\alpha\left[ \bar{x}_\alpha^I(0) -
\bar{x}_\alpha^{II}(0)\right] .
\label{z2w}
\end{equation}
Since the gauge is assumed to be smooth, the evolution of the charge
centers must also be smooth. Being defined in this way, $\Delta$
is clearly a mod-2 quantity, and as shown in Ref.~\onlinecite{Fu-PRB06}
it represents the desired $\z2$ invariant.

However, if the gauge breaks $\TR$ symmetry or it is not continuous
in the half-cycle, Eq.~(\ref{z2w}) no longer defines a topological
invariant. A discontinuity in the gauge in the process of
the half cycle can change $\Delta$ by $1$, so the mod-2 property is
lost. Breaking $\TR$ in the gauge choice means that
the corresponding centers are not
necessarily degenerate at $t=0$ and $t=\TT/2$.  In fact,
$\Delta$ can even take non-integer values in this 
case.\cite{Soluyanov-PRB11}

The above argument implies that in order to compute the $\z2$ invariant
via Eq.~(\ref{z2w}), one needs a gauge that satisfies both
$\TR$-invariance and continuity on the half-torus. We now argue that the
gauge that corresponds to 1D maximally localized WFs at each $t$ has
the desired properties,
as long as these WFs are chosen to evolve
smoothly as a function of $t$.
The criterion introduced in
Ref.~\onlinecite{Marzari-PRB97} for constructing the maximally
localized WFs was that the
gauge choice should provide the minimum possible quadratic spread
$\Omega=\sum_n[\langle 0n | \r^2 |0n \rangle - \langle 0n | \r |0n
\rangle^2]$. In 1D, the maximally localized WFs constructed
according to this criterion are eigenstates of the position
operator $\hat{X}$ in the band subspace.\cite{Marzari-PRB97, Kivelson-PRB82}
Since this operator commutes with $\theta$, its eigenvalues will
be doubly degenerate and its eigenstates will come in Kramers
pairs at $t=0$ and $\TT/2$.

To prove continuity of this gauge in $k$, let us briefly discuss how
to enforce it on a $k$-mesh $k_{j+1}=k_j+\Delta k$ by carrying 
out a multi-band parallel-transport
construction along the Brillouin zone.\cite{Marzari-PRB97}
At a given value of $t$, starting from $k=0$ one constructs overlap matrices
$M^{(k_j,k_{j+1})}_{mn}=\langle u_{mk_j}| u_{n k_{j+1}} \rangle$
in such a way that they are Hermitian. This can be done in a unique way
by means of the singular value decomposition
$M=V\Sigma W^\dagger$, where $\Sigma$ is positive real diagonal and
$V$ and $W$ are unitary matrices. With this decomposition a unitary rotation
of the states at $k_{j+1}$ by $WV^\dagger$ leaves $M^{(k_j,k_{j+1})}$ Hermitian.
Repeating this procedure, one finds that states $|\psi_{nk}\rangle$
at $k=2\pi$ are related to those at $k=0$ by a unitary rotation
$\Lambda$, whose eigenvalues $\lambda_n=e^{-i\bar{x}_n}$
give the 1D maximally-localized WCCs $\bar{x}_n$.
The corresponding eigenvectors can be used to define a gauge
that is continuous in $k$ for a given value of $t$.
The continuity vs.~$t$ on the half-torus is achieved by tracing
the evolution of the WCCs $\bar{x}_n$ as a function of $t$, with
the $n$'th state of the gauge constructed from the eigenvectors
associated with the $n$'th smoothly evolving WCC $\bar{x}_n(t)$.

Having established a particular gauge choice in which
Eqs.~(\ref{z2}) and (\ref{z2w}) are valid, it is straightforward
in principle to obtain the $\z2$ invariant.
Indeed, Eq.~(\ref{z2w}) implies that the $\z2$ invariant can be
determined simply by testing whether the WCCs change partners when
tracked continuously from $t$=0 to $T/2$.   This is the essence
of our approach.  We stress that no explicit construction of a
smooth gauge on the half-torus is necessary; we simply
track the evolution of the WCCs on the half-torus.

In practice,
when working on a discrete mesh of $t$ values when many bands
are present, it may not be entirely straightforward to enforce
the continuity with respect to $t$.  In the next section we
present a simple and automatic numerical procedure that is robust
in this respect, and use it to illustrate the calculation of the
$\z2$ invariants for several materials of interest.

\section{\label{Sec:3} Numerical Implementation}

The method outlined above, in which the WCCs obtained with the 1D
maximally-localized gauge are used to compute the $\z2$ invariant
via Eq.~(\ref{z2w}), can be implemented by plotting the
WCCs at each point on the $t$ mesh and then visually tracking the
evolution of each WCC, as we describe next in Sec.~\ref{sec:sec3a}.
However, we find that a more straightforward and more easily
automated approach is to track the {\it largest gap} in the
spectrum of WCCs instead.  This gives rise to our proposed method,
which is described in Sec.~\ref{sec:sec3b}.

\subsection{\label{sec:sec3a} Tracking WCC locations}

\begin{figure}
\begin{center}
\includegraphics[width=3.4in]{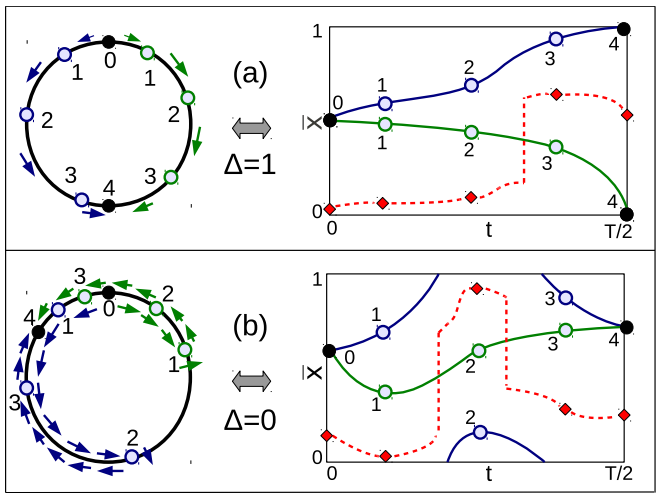}
\end{center}
\caption{Sketch of evolution of Wannier charge centers
(WCCs) $\bar{x}$ vs.\ time $t$ during an adiabatic pumping process.
Regarding $\bar{x}\in[0,1]$ as a unit circle and $t\in[0,\TT/2]$
as a line segment, the cylindrical $(\bar{x},t)$ manifold is
represented via a sequence of circular cross sections at left,
or as an unwrapped cylinder at right.  Each red rhombus marks the
middle of the largest gap between WCCs at given $t$. (a) $\z2$
insulator; WCCs wind around the cylinder.  (b) Normal
insulator; WCCs reconnect without wrapping the cylinder.}
\label{fig:fig1}
\end{figure}

Let us first interpret Eq.~(\ref{z2w}) in terms of the winding of
the WCCs around the BZ during the half-cycle $t\in[0,\TT/2]$.
Since the WCCs are defined modulo 1, one can imagine the
$\bar{x}_n$ living on a circle of unit circumference, as
illustrated in the left panels of Fig.~\ref{fig:fig1}.  During the
pumping process, the WCCs migrate along this circle.  The system
will be in the $\z2$-odd state ($\delta$=1) if and only if the
WCCs reconnect after the half cycle in such a way as to
wrap the unit circle an odd number of times.

Consider, for example, the case of only two occupied bands,
as sketched in Fig.~\ref{fig:fig1}.  The top panel shows the
$\z2$-odd case; the blue and green arrows show the evolution of the
first and second WCC from $t_0$ (=\,0) to $t_4$ (=\,$\TT/2$),
and they meet in such a way that the unit circle is wrapped
exactly once.
Correspondingly, as shown in the right-hand part of the figure,
the WCCs ``exchange partners'' during the pumping process
(i.e., two bands belonging to the same Kramers pair
at $t=0$ do not rejoin at $t=T/2$).\cite{Fu-PRB06} 
For the $\z2$-even case shown in the bottom panel,
by contrast, the unit circle is wrapped zero times, and no such
exchange of partners occurs.

If one has access to the continuous evolution of the WCCs vs.~$t$,
as shown by the solid blue and green curves in Fig.~\ref{fig:fig1},
this method works in principle for an arbitrary
number of occupied bands (i.e., WFs per unit cell).
An illustrative example with many bands appears in Fig.~(1) of
Ref.~\onlinecite{Ringel-arx10}.  Either the ``bands'' $\bar{x}_n$
exchange partners in going from $t=0$ to $t=T/2$ ($\phi=0$ to
$\phi=\pi$ in their notation), or they do not, implying $\z2$
odd or even respectively.
Equivalently, one can draw an arbitrary
continuous curve starting within a gap at $t=0$ and ending within a
gap at $t=T/2$; the system is $\z2$-odd if this curve crosses the
WCC bands an odd number of times, or $\z2$-even otherwise.

In practice, however, one will typically have the WCC values only
on a discrete mesh of $t$ points, in which case the connectivity
can be far from obvious.  Certainly one cannot simply make
the arbitrary branch cut choice $\bar{x}_n\in[0,1]$, sort the
$\bar{x}_n$ in increasing order, and use the resulting indices to
define the paths of the WCCs.  This would, for example, give an
incorrect evolution from $t_1$ to $t_2$ in Fig.~\ref{fig:fig1}(b),
since one WCC passes through the branch cut in this interval,
apparently jumping discontinuously from the ``top'' to the
``bottom'' of the unwrapped cylinder at right.  (A similar jump
happens again near $t_3$.)

One possible approach is that of Ref.~\onlinecite{Ringel-arx10}
mentioned above, i.e., to increase the $t$ mesh density until,
by visual inspection, the connectivity becomes obvious.
However, this becomes prohibitively expensive in the
first-principles context, since a calculation of many (typically
10-30) bands would have to be done on an extremely fine mesh
of $t$ points.  It is typical for some of the WCCs to cluster
rather closely together during part of the evolution in $t$;
if this clustering happens near the artificial branch cut, it can
become very difficult to determine the connectivity from one $t$
to the next, even if a rather dense mesh of $t$ values is used.
Moreover, an algorithm of this kind is difficult to automate.
For these reasons, we find that the direct approach of plotting
the evolution of the WCCs is not a very satisfactory algorithm
for obtaining the topological indices, at least in the case of
a large number of occupied bands.

\subsection{\label{sec:sec3b} Tracking gaps in the WCC spectrum}
\label{sec:gaps}

Here we propose a simple procedure that overcomes the above
obstacles, allowing the $\z2$ invariant to be computed in a
straightforward fashion.  The main idea is to concentrate on the
{\it largest gap between WCCs}, instead of on the individual WCCs
themselves.
As explained above and illustrated by the red dashed curve in
Fig.~\ref{fig:fig1}, the path following the largest gap in
$\bar{x}_n$ values (with vertical excursions
at critical values of $t$) crosses the $\bar{x}_n$ bands
a number of times that is equal, mod 2, to the $\z2$ invariant.
Our approach, in which we choose this path as an especially suitable
one for discretizing,
can be implemented without reference
to any branch cut in the determination of the $\bar{x}_n$, allowing
the $\z2$ invariant to be determined from the flow of WCCs on
the cylindrical $(\bar{x},t)$ manifold directly.

As in Fig.~\ref{fig:fig1}, we again consider a set of $M$
circular sections of the cylinder that correspond to the pumping
parameter values $t^{(m)}=T(m-1)/2M$, where $m\in[0,M]$. At each
$t_m$ we define $z^{(m)}$ to be the center of the largest gap
between two adjacent WCCs on the circle. (If two gaps are of
equal size, either can be chosen arbitrarily.)  For definiteness
we choose $z^{(m)}\in[0,1)$, but as we shall see shortly, the
branch choice is immaterial.  In the continuous limit
$M\rightarrow\infty$, $z(t)$ takes the form of a series of path
segments on the surface of the cylinder, with discontinuous
jumps in the $\bar{x}$ direction at certain critical parameter
values $t_j$.  Our algorithm consists in counting the number
of WCCs jumped over at each $t_j$, and summing them all mod 2.
As becomes clear from an inspection of Fig.~\ref{fig:fig1} and
similar examples of increasing complexity, the WCCs exchange
partners during the evolution from $t$=0 to $\TT/2$ only if this
sum is odd, so that this sum determines the $\z2$ invariant
of the system.

The approach generalizes easily to the case of discrete $z^{(m)}$.
Let $\Delta_m$ be the number of WCCs $\bar{x}_n^{(m+1)}$ that
appear between gap centers $z^{(m)}$ and $z^{(m+1)}$, mod 2.
As we shall see below, this can be computed in a manner that is
independent of the branch cut choices used to determine
the $\bar{x}_n^m$ and $z^{(m)}$.  Then the overall $\z2$
invariant is just
\begin{equation}
\Delta=\sum_{m=0}^M\Delta_m\quad \hbox{mod}\;2.
\label{Delta}
\end{equation}

This argument is illustrated in the right-hand panels of
Fig.~\ref{fig:fig1} for the two band-case and $M=4$. The rectangles
represent the surface of the
cylinder in the parameter space, and should be regarded as glued
along the longer sides. The circles correspond to $\bar{x}_n^{(m)}$
values, while each red rhombus represents the center $z^{(m)}$
of the largest gap between $\bar{x}_n^{(m)}$ values.  In
Fig.~\ref{fig:fig1}(a) there is one jump that occurs between
$m$=2 and $m$=3, in which one WCC is jumped over; thus,
$\Delta_m=0$ except for $\Delta_2=1$, giving
$\Delta$=1.  In Fig.~\ref{fig:fig1}(b), on the other hand, there
are two jumps, once between $m$=1 and $m$=2 and again
between $m$=2 and $m$=3, so that $\Delta_1=\Delta_2=1$ and
$\Delta=0\hbox{ (mod 2)}$.

\begin{figure}
\begin{center}
\includegraphics[width=2.6in, clip=true]{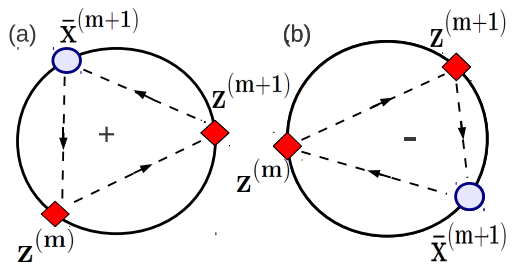}
\end{center}
\caption{Sketch illustrating the method used to determine whether
$\bar{x}_n^{(m+1)}$ lies between $z^{(m)}$ and $z^{(m+1)}$ in
the counterclockwise sense when mapped onto the complex unit circle.
(a) Yes, since the directed area of the triangle is positive.
(b) No, since it is negative.}
\label{fig:circle}
\end{figure}

We now show how the $\Delta_m$ can be computed straightforwardly
in a manner that is insensitive to the branch-cut choices made in
determining the $\bar{x}_n^m$ and $z^{(m)}$.  We use the fact that
the directed area of a triangle defined by angles $\phi_1$,
$\phi_2$, and $\phi_3$ on the unit circle is
\footnote{This follows from the fact that the directed area of the
  triangle defined by vertices $z_j$ in the complex plane is
  $\mathrm{Im}[z_1^*z_2+z_2^*z_3+z_3^*z_1]$; specializing this to
  $z_j=\exp(i\phi_j)$ yields Eq.~(\ref{darea}).}
\begin{equation}
g(\phi_1,\phi_2,\phi_3)=
\sin(\phi_2-\phi_1)+\sin(\phi_3-\phi_2)+\sin(\phi_1-\phi_3).
\label{darea}
\end{equation}
Therefore the sign of
$g(\phi_1,\phi_2,\phi_3)$ tells us whether or not $\phi_3$ lies
``between'' $\phi_1$ and $\phi_2$ in the sense of counterclockwise
rotation.  Identifying $\phi_1=2\pi z^{(m)}$, $\phi_2=2\pi z^{(m+1)}$ and
$\phi_3=2\pi \bar{x}_n^{(m+1)}$, as in Fig.~\ref{fig:circle},
it follows that
\begin{equation}
(-1)^{\Delta_m}=\prod_{n=1}^{\cal N} \mathop{\rm sgn}\left[
g(2\pi z^{(m)},2\pi z^{(m+1)}, 2\pi\bar{x}_n^{(m+1)}) \right],
\label{formula}
\end{equation}
where $\mathop{\rm sgn}(x)$ is the sign function.  The $\Delta_m$
defined in this way is precisely the needed count of WCCs jumped
over, mod 2, in evolving from $m$ to $m+1$.

As a last detail, we discuss the case of possible degeneracies
between the three arguments of $g(\phi_1,\phi_2,\phi_3)$.
First, note that
$z^{(m+1)}=\bar{x}_n^{(m+1)}$ is impossible, since $z^{(m+1)}$ is
by definition in a gap between $\bar{x}_n^{(m+1)}$ values.
If the mesh spacing in $t$ is fine enough, then by continuity we
expect that $z^{(m)}=\bar{x}_n^{(m+1)}$ will also be unlikely.
It is recommended to test whether these values ever approach
within a threshold distance, and restart the algorithm with a
finer $t$ mesh if such a case is encountered; two cases of this
kind are discussed later in Sec.~\ref{sec:applic}.  Finally,
it can happen that $z^{(m)}=z^{(m+1)}$.  In this case, the signum
function (which technically assigns value 0 to argument 0) should
be replaced in Eq.~(\ref{formula}) by a function that returns
$s$ whenever $z^{(m)}=z^{(m+1)}$, where $s$ is chosen once and for
all to be either +1 or $-1$.  Since the same degeneracy appears
in every term of the product over $\cal N$ factors in
Eq.~(\ref{formula}), where $\cal N$ is even, the choice of $s$ is
arbitrary as long as it is applied consistently.

The above-described algorithm, based on
Eqs.~(\ref{Delta}-\ref{formula}), constitutes one of the principal
results of the present work.  The implementation of this
algorithm is straightforward, and allows for an efficient and
robust determination of the $\z2$ invariant even when
many bands are present, and even for only moderately fine mesh
spacings.  In Sec.~\ref{sec:applic}, we will demonstrate the
successful application of this approach to the calculation of
the strong and weak topological indices of some real materials.

\subsection{Application to 2D and 3D $\TR$-invariant insulators}

As pointed out in Ref.~(\onlinecite{Fu-PRB06}), the pumping
process discussed above for a 1D system is the direct analogue of
a 2D $\TR$-invariant insulator, i.e., one whose Hamiltonian is
subject to the condition $H(-\k)=\theta^{-1}H(\k)\theta$.  To see
this, let $\k=\sum_i k_i{\bf b}_i/2\pi$, where ${\bf b}_1$ and
${\bf b}_2$ have been chosen as primitive reciprocal lattice vectors.
Then we can let
$k_1$ and $k_2$ play the roles of $k$ and $t$ respectively.
Just as $H(k,t)$ displays $\TR$ symmetry of $H(x)$ at $t$=0 and
$\TT/2$,
so $H(k_1,k_2)$, regarded as the Hamiltonian $H(x_1)$ of
a fictitious 1D system for given $k_2$, 
is $\TR$-invariant at $k_2=0$ and $\pi$.  
The Wannier functions of the effective 1D
system can be understood as ``hybrid Wannier functions'' that
have been Fourier transformed from $k$ space to $r$ space only
in direction 1, while remaining extended in direction 2.
The topological $\z2$ invariant of the 2D system can therefore be
determined straightforwardly by applying the approach outlined
above.

A topological phase of a 3D $\TR$-symmetric insulator is described by
one strong topological index $\nu_0$ and three weak indices
$\nu_1$, $\nu_2$, and $\nu_3$.\cite{Moore-PRB07, Roy-PRB09-b, Fu-PRL07}
These indices may be understood as follows. Again letting
$\k=\sum_i k_i{\bf b}_i/2\pi$, there are eight $\TR$-invariant points
$\Gamma_{(n_1,n_2,n_3)}$, where $n_i=0$ or 1 denotes $k_i=0$ or $\pi$
respectively.  These eight points may be thought of as the vertices
of a parallelepiped in reciprocal space whose six faces are labeled by
$n_1$=0, $n_2$=0, $n_3$=0, $n_1$=1, $n_2$=1, and $n_3$=1.
On any one of these six faces, the Hamiltonian $H(\k)$, regarded
as a function of two $k$ variables, can be thought of as the
Hamiltonian of a fictitious 2D $\TR$-symmetric system, and the argument
of the previous paragraph can thus be applied to each of these
six faces separately. The three weak indices $\nu_{i=1,2,3}$ are
defined to be the $\z2$ invariants associated with the three surfaces
$n_1$=1, $n_2$=1, and $n_3$=1.\cite{Fu-PRL07} These weak indices
obviously depend on the choice of reciprocal lattice vectors.
The strong index $\nu_0$ is the sum (mod 2) of the $\z2$ invariants
of the $n_j$=0 and $n_j$=1 faces for any one of the $j$ (implying
some redundancy among the six indices); it is also a $\z2$
quantity, but is independent of the choice of reciprocal lattice
vectors.\cite{Fu-PRL07,Moore-PRB07}

Thus, a complete topological
classification in 3D, given by the index $\nu_0;(\nu_1 \nu_2 \nu_3)$,
can be obtained by applying our analysis to each of these six faces in the
3D Brillouin zone.
Note that in general, this determines the strong index $\nu_0$ with
some redundancy, providing a check on the internal consistency of
the method.  However, symmetry considerations often play a role.
For systems having a 3-fold symmetry axis, for example, one
typically needs to compute the $\z2$ index on only two faces,
as we shall see below.

\section{Application to real materials}
\label{sec:applic}

In this section we discuss the application of the above-described
method to real materials. First, we illustrate the validity of the
approach for centrosymmetric Bi and Bi$_2$Se$_3$, where weak and
strong indices may alternatively be computed directly from the
parities of the occupied
Kramers pairs at the eight $\TR$-invariant momenta.\cite{Fu-PRB07}
We then apply the method to noncentrosymmetric crystals of
GeTe and strained HgTe, showing that the first is a trivial insulator,
while the latter is a strong topological insulator under both
positive and negative strains along $[001]$ and under positive strain
along $[111]$.

The calculations were carried out in the framework
of density-functional theory\cite{Kohn-PR65} using the local-density
approximation with the exchange and correlation
parametrized as in Ref.~\onlinecite{Goedecker-PRB96}.  We used HGH
pseudopotentials\cite{HGH-PRB98} with semicore $5d$-states included
for Hg, while for all other elements only the $s$ and $p$ valence
electrons were explicitly included.
The calculations were carried out using the ABINIT code
package\cite{Gonze-ZK05,Gonze-CPC09} with a $10\times10\times10$ $\k$-mesh
for the self-consistent field calculations and a $140$\,Ry planewave
cutoff. The spin-orbit interaction was included in the calculation
via the HGH pseudopotentials.
Note that the overlap matrices $M^{(k_j,k_{j+1})}_{mn}$ defined in
Sec.~\ref{sec:formulation}, are the same as those needed for the
calculation of the electric polarization\cite{King-Smith-PRB93}
or the construction of maximally-localized Wannier
functions,\cite{Marzari-PRB97} and are thus readily available in
many standard {\it ab initio} code packages including ABINIT.

\subsection{Centrosymmetric materials}

We start by illustrating the method with the examples of Bi and
Bi$_2$Se$_3$. Although Bi is a semimetal, its ten lowest-lying
valence bands are separated from higher ones by an energy gap
everywhere in the BZ, so in this
case the topological indices describe the topological character of
a particular group of bands.  Since this is not the occupied subspace
of an insulator, these topological indices are not ``physical,''
but it is still of interest to compute them and compare with
methods based  on the parity eigenvalues.\cite{Fu-PRB07}  According
to the latter approach, the group of ten lowest-lying bands of Bi
was shown to be topologically trivial.\cite{Fu-PRB07}
Bi$_2$Se$_3$, on the other hand, is a true insulator, and the parity
approach demonstrated that it is a strong topological
insulator.\cite{Zhang-NatPhys09}

Bi and Bi$_2$Se$_3$ both belong to the rhombohedral space group
$R\bar{3}m$ (\#166), which has a 3-fold rotational axis. Thus, it is
enough to compute only one weak $\z2$ index, say for $n_1=1$, since all
three of them are equal by symmetry.  To get the strong index, one
just needs to compute just one more of the $\z2$ invariants, say for
$n_1$=0.

\begin{figure}
\begin{center}
\includegraphics[width=2.8in, bb=2 27 340 384]{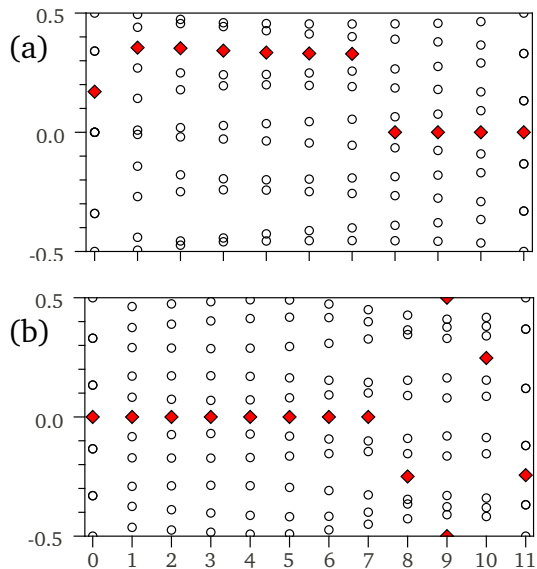}
\end{center}
\caption{Evolution of Bi WCCs $\bar{x}_n$ (circles) in the
$r_3$ direction vs.\ $k_2$ at (a) $k_1$=0; (b) $k_1$=$\pi$.
Red rhombus marks midpoint of largest gap.  $k_2$ is sampled
in ten equal increments from 0 to $\pi$, except that an
extra point is inserted midway in the last segment in panel (b)
(see text).}
\label{fig:fig2}
\end{figure}

Our results for Bi, obtained with the lattice parameters used in
previous studies,\cite{Gonze-PRB90} are presented in Fig.~\ref{fig:fig2}.
Panels (a) and (b) show the determination of the $\z2$ invariant at
$n_1$=0 and $n_1$=1 respectively, with $k_2$ treated as the pumping
parameter (like $t$) for an effective 1D system with wavevector
$k_3$.  The $k_2$ axis was initially discretized into ten equal
intervals ($m=1,...,10$) running from 0 to $\pi$, but for
reasons discussed below an extra point (number 10 on the horizontal
axis of the plot) was inserted midway in the last segment to make
a total of eleven $m$ values in Panel (b).  As noted above,
we are treating a group of ten valence bands labeled by $n$, so
we have an array of WCC values $\bar{x}_n^{(m)}$ whose values are
indicated by the black circles in the plot.  These form
Kramers pairs at $k_2$=0 and $\pi$, but not elsewhere.  Each red
rhombus indicates the center $z^{(m)}$ of the largest gap between
adjacent $\bar{x}_n^{(m)}$ values, as discussed in Sec.~\ref{Sec:3}.

Looking first at Fig.~\ref{fig:fig2}(a), we see that the gap center jumps
over one WCC at $m$=1, and then over three WCCs at $m$=7, for a total
of four, which is even.  In Fig.~\ref{fig:fig2}(b) we get a total of
$2+7+3+4=16$ jumps, which is again even.  The visual determinations
of the number of jumped bands is confirmed by the application of the
automated procedure of Eqs.~(\ref{Delta}-\ref{formula}).  Thus,
both $\z2$ indices are $0$, and the 3D index is $0;(000)$, indicating
a normal band topology as anticipated.\cite{Fu-PRB07,Fukui-JPSJ07}

We now discuss the above-mentioned insertion of one extra $k_2$
point in Fig.~\ref{fig:fig2}(b).  This was necessary because
the gap center $z^{(9)}$ at $k_2=0.9\pi$ had almost the same
value as one of the WCC values at $k_2=\pi$ (now labeled as `11'
on the horizontal axis), making it ambiguous
whether or not that $x_n$ value should be counted as one of
the ones that has been jumped over.  To resolve this difficulty,
we included an extra step at $k_2=0.95\pi$ (now labeled as `10'
on the horizontal axis).  The reason for the fast motion of
the WCC in this case is that the minimum gap to the next higher
(eleventh) band becomes rather small near $k_2=\pi$.

Note that the detection of this kind of problem does not have to
be done by visual inspection, but can be automated in the context
of Eqs.~(\ref{Delta}-\ref{formula}).  As already mentioned in
Sec.~\ref{sec:gaps}, we simply test whether any $\bar{x}_n^{(m+1)}$
approaches within a certain threshold of $z^{(m)}$ (mod 1); if so,
we flag the interval in question for replacement by a finer mesh.
Still, it is recommended to choose a mesh that is fine enough
so that this threshold is rarely encountered, with a finer mesh
recommended in cases where the minimum band gap is small.
\footnote{In the vicinity of a small
  gap, it is also advisable to reduce the mesh spacing along the
  $k$-point strings used for the parallel transport construction.}

\begin{figure}
\begin{center}
\includegraphics[width=2.8in, bb=2 0 345 365]{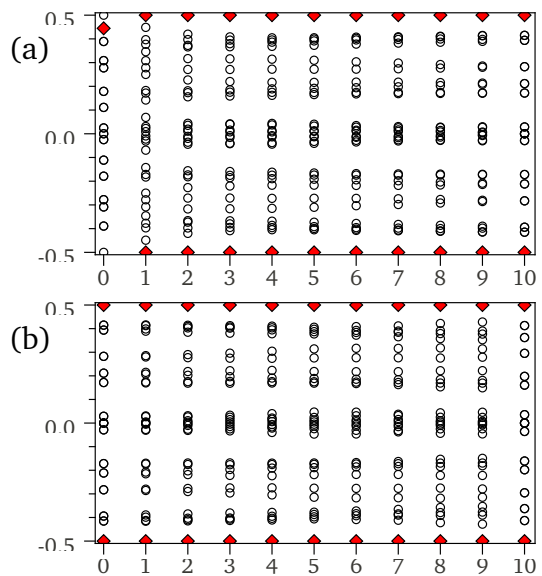}
\end{center}
\caption{Evolution of Bi$_2$Se$_3$ WCCs $\bar{x}_n$ (circles) in the
$r_3$ direction vs.\ $k_2$ at (a) $k_1$=0; (b) $k_1$=$\pi$.
Red rhombus marks midpoint of largest gap.  $k_2$ is sampled
in ten equal increments from 0 to $\pi$.}
\label{fig:fig3}
\end{figure}

The analysis of the same $n_1$=0 and $n_1$=1 faces for the 28
WCCs of Bi$_2$Se$_3$ is illustrated in Fig.~\ref{fig:fig3}.
The experimental lattice parameters\cite{Wiese-JPCS60}
were used. Here there are no jumps over WCCs except for a single
one in the very first step in the top panel ($n_1$=0).  It follows
that the topological index is $1;(000)$, in accord with previous
studies.\cite{Zhang-NatPhys09}

\subsection{Noncentrosymmetric materials}

We now proceed to systems without inversion symmetry, which are
the principal targets of our method since an analysis based on parity
eigenvalues is not possible.

\begin{figure}
\begin{center}
\includegraphics[width=2.8in, bb=2 0 345 365]{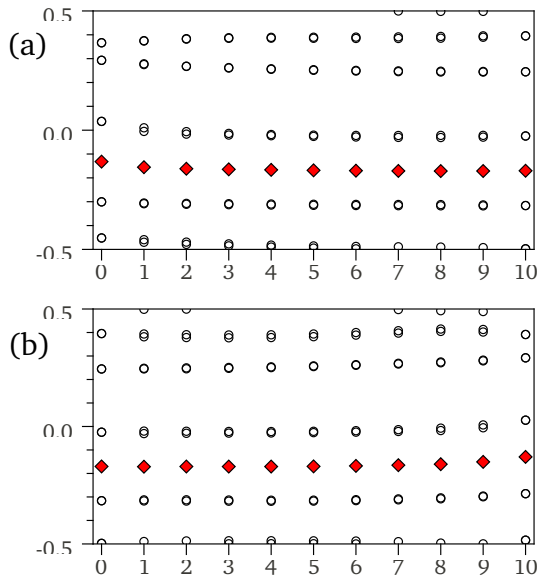}
\end{center}
\caption{Evolution of GeTe WCCs $\bar{x}_n$ (circles) in the
$r_3$ direction vs.\ $k_2$ at (a) $k_1$=0; (b) $k_1$=$\pi$.
Red rhombus marks midpoint of largest gap.  $k_2$ is sampled
in ten equal increments from 0 to $\pi$.}
\label{fig:fig4}
\end{figure}

GeTe belongs to the rhombohedral $R3m$ space group ($\#160$) and has no
inversion symmetry, although like Bi and Bi$_2$Se$_3$ it has a 3-fold
rotational symmetry, so that only two
reciprocal-space faces have to be studied.
The experimental lattice parameters\cite{Onodera-PRB97} were used, and
the evolution of the 10 WCCs is presented in Fig.~\ref{fig:fig4}
following similar conventions as for Bi and  Bi$_2$Se$_3$.
For both faces Eq.~(\ref{formula}) gives a trivial $\z2$
index, with the center of the largest gap making no jumps, so that GeTe
is in the topologically trivial state $0;(000)$. This result could have
been anticipated from the fact that the spin-orbit interaction in GeTe
is weak, as reflected in the approximate pairwise degeneracy of the
WCCs throughout the evolution.

Finally, let us consider the more interesting case of strained HgTe.
In the absence of strain this is a zero-band-gap material. Any
anisotropic strain breaks the four-fold symmetry at $\Gamma$, making
it possible that the gap might open. Based on an adiabatic continuity
argument, HgTe was predicted to be a strong topological insulator
under compressive strain in the $[001]$ direction.\cite{Fu-PRB07} This
was later verified with tight-binding calculations.\cite{Dai-PRB08,
Bernevig-Science06}  Application of our approach to HgTe under
uniaxial strain also confirms that HgTe is a strong topological
insulator, with index $1;(000)$, under both positive and
negative\cite{Fu-PRB07} 2\% strains along the $[001]$
direction (not shown).  This means that although the positive-strain
and negative-strain states are separated by a gap closure at zero
strain, there is no topological phase transition associated with
this gap closure.

\begin{figure}
\begin{center}
\includegraphics[width=3.4in, bb=2 2 480 282]{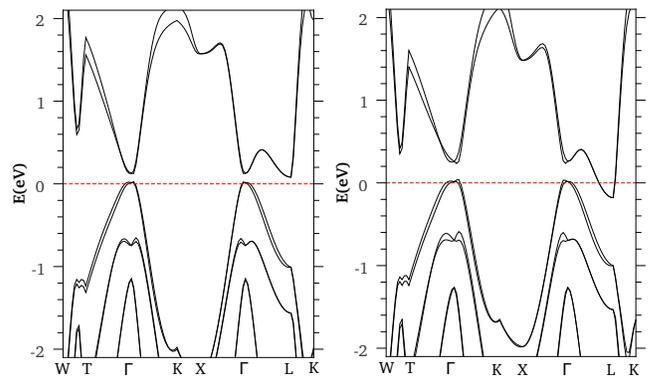}
\end{center}
\caption{Band structure along high-symmetry lines of the
undistorted FCC structure for HgTe under tensile strain in
the $[111]$ direction. (a) $+2\%$ strain.  (b) $+5\%$ strain.}
\label{fig:bs}
\end{figure}

We also studied strains in the $[111]$ direction. Under compressive
strains of $-2\%$ and $-5\%$ the system becomes metallic and the
direct band gap vanishes, so that no topological index can be
associated with the occupied space. Under tensile strain of $+2\%$
we find that HgTe becomes a narrow-gap semiconductor with an indirect
energy gap of $E_g=0.054$\,eV, while for $+5\%$ strain it becomes
metallic.  Even at $+5\%$ strain, however, the lowest 18 bands remain
separated from higher ones by an energy gap at all $\k$, so that, as for
Bi, one can still assign a topological index to this isolated
group of bands. The computed band structures for both cases are
illustrated in Fig.~\ref{fig:bs} along lines connecting the high
symmetry points of the undistorted FCC structure.

\begin{figure}
\begin{center}
\includegraphics[width=2.8in, bb=0 0 346 360]{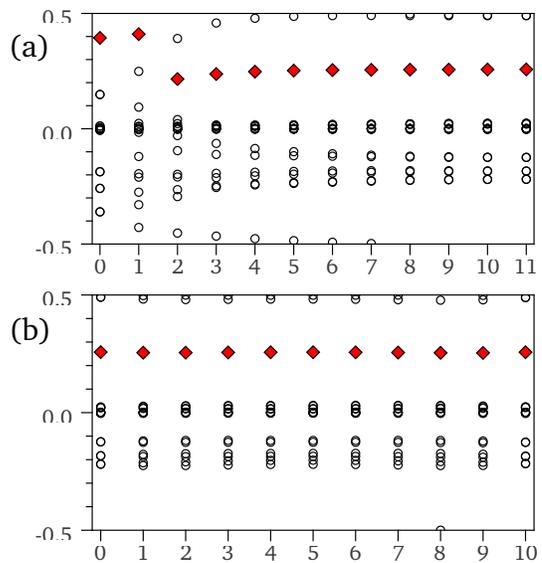}
\end{center}
\caption{Evolution of WCCs for HgTe under +2\% strain in the
[111] direction.  WCCs $\bar{x}_n$ (circles) in the
$r_3$ direction are plotted vs.\ $k_2$ at (a) $k_1$=0; (b) $k_1$=$\pi$.
Red rhombus marks midpoint of largest gap.  $k_2$ is sampled
in ten equal increments from 0 to $\pi$, except that an
extra point is inserted midway in the first segment in panel (a)
(see text).}
\label{fig:fig5}
\end{figure}

The space group of $[111]$-strained HgTe is rhombohedral
$R3m$ ($\#166$), the same as for GeTe, so that again only two
$\z2$ indices need to be calculated.  The results of our
WCC analysis for the case of $+2\%$ strain are shown in
Fig.~\ref{fig:fig5}.
We find $\z2$=1 and $\z2$=0 for $n_1$=0 and $n_2$=1 respectively,
so that the topological class is $1;(000)$.  The behavior in
Panel (b) is rather uninteresting, since the gap is large everywhere
on the $n_2$=1 face.  However, in Panel (a) we again find an
example of a rapid change of WCCs with $k_2$, which was repaired
by inserting an extra point (the one now labeled `1' on the
horizontal axis) at $k_2=0.05\pi$.  Actually, we anticipated the
need for this denser sampling for small $k_2$ from the fact
that the zero-strain gap closure occurs at $\Gamma$, so that a
delicate dependence on $\k$ near the BZ center was expected.

\section{Summary and conclusions}
\label{sec:concl}

We have proposed a new approach for calculating
topological invariants in $\TR$-invariant systems. The method is based
on following the evolution of hybrid Wannier charge centers, and is very
general, being easily applicable in both tight-binding and DFT contexts.
The needed ingredients are the same as those needed for the
calculation of the electric polarization or the construction
of maximally-localized Wannier functions, and are thus readily
available in standard code packages.
The present algorithm is relatively
inexpensive, however, because the analysis is confined to a small
number of 2D slices of the 3D Brillouin zone.  The method is
easily automated and remains robust even when many bands are present.
We hope that our method can help to
make the search for topological phases in noncentrosymmetric
materials a routine task, and that it will lead to further progress
in this rapidly developing field.

Note: In the final stages of preparing this manuscript, we became aware
of independent work by Yu {\it et al.}\cite{Yu-arx11} that is
closely related.  These authors carry out a similar analysis based
on WCCs, but without the automated analysis described in our
Sec.~\ref{Sec:3}.

\section{ACKNOWLEDGMENTS}

This work was supported by NSF Grant DMR-1005838.  We thank D. R.
Hamann for useful discussions.

\bibliography{paper}
\end{document}